\documentclass[review]{elsarticle}

\usepackage{lineno,hyperref}

\usepackage{epsfig}

\modulolinenumbers

\journal{Journal of \LaTeX\ Templates}

\begin{document}

\begin{frontmatter}

\title{A model of irreversible jam formation in dense traffic}

\author[label1,label2]{J. G. Brankov}
\author[label1,label2]{N. Zh. Bunzarova}
\address[label1]{Bogoliubov Laboratory of Theoretical Physics, Joint Institute for Nuclear Research, 141980 Dubna, Russia}
\address[label2]{Institute of Mechanics, Bulgarian Academy of Sciences, 1113 Sofia, Bulgaria\corref{mycorrespondingautho}}
\cortext[N. Zh. Bunzarova]{Corresponding author}
\ead{nadezhda@imbm.bas.bg}

\author[label2]{ N. C. Pesheva}

\author[label1]{V. B. Priezzhev}

\begin{abstract}

We study an one-dimensional stochastic model of vehicular traffic on open segments of a single-lane road of finite size $L$. The vehicles obey a stochastic discrete-time dynamics which is a limiting case of the generalized Totally Asymmetric Simple Exclusion Process. This dynamics has been previously used by Bunzarova and Pesheva [Phys. Rev. E 95, 052105 (2017)] for an one-dimensional model of irreversible aggregation. The model was shown to have three stationary phases: a many-particle one, MP, a phase with completely filled configuration, CF, and a boundary perturbed MP+CF phase, depending on the values of the particle injection ($\alpha$), ejection ($\beta$) and hopping ($p$) probabilities.

Here we extend the results for the stationary properties of the MP+CF phase, by deriving exact expressions for the local density at the first site of the chain and the probability P(1) of a completely jammed configuration. The unusual phase transition, characterized by jumps in both the bulk density and the current (in the thermodynamic limit), as $\alpha$ crosses the boundary $\alpha =p$ from the MP to the CF phase, is explained by the finite-size behavior of P(1). By using a random walk theory, we find that, when $\alpha$ approaches from below the boundary $\alpha =p$, three different regimes appear, as the size $L\rightarrow \infty$: (i) the lifetime of the gap between the rightmost clusters is of the order $O(L)$ in the MP phase; (ii) small jams, separated by gaps with lifetime $O(1)$, exist in the MP+CF phase close to the left chain boundary; and (iii) when $\beta =p$, the jams are divided by gaps with lifetime of the order $O(L^{1/2})$. These results are supported by extensive Monte Carlo calculations.

\end{abstract}

\begin{keyword}

Non-equilibrium phenomena, One-dimensional stochastic process, Stationary states, Congested traffic, Irreversible jam formation

\end{keyword}

\end{frontmatter}


\section{Introduction}

Our aim here is to study a discrete-time discrete-space model of congested traffic on finite  segments of a single-lane road, which allows a detailed description of the growth of clusters of jammed vehicles under different boundary conditions. We assume that the vehicles (particles) obey a stochastic dynamics with the following properties: (a) Existing clusters of jammed vehicles do not break into parts during their motion along the selected segment of the road; (b) With given probability, such clusters are translated as a whole entity one site to the right, provided the target site is empty, in the same way as isolated vehicles do; (c) Any two vehicles or clusters of vehicles, occupying consecutive positions on the chain, may become nearest-neighbors and merge irreversibly into a single cluster. Here we note that property (b) is characteristic of the synchronized vehicle movement in high density traffic. Due to the property (c), the size of the moving jam grows monotonically with time until it spreads on the whole road segment, or leaves the system before that happens. Properties (a) and (b) correspond to the case when the drivers in a jam closely follow the vehicles in front of them, thus keeping the minimum safe distance between the vehicles. This is a simplified picture of synchronously moving vehicles in a jam.

It is the balance between the typical time interval separating the appearance of gaps at the entrance of the road, and the average portion of time that a gap between consecutive clusters persists, that determines the stationary probability P(1) of a completely jammed configuration of the system. This dynamics has already been used in a one-dimensional model of irreversible aggregation \cite{BP16}. The model was shown to have three stationary phases: a many-particle one, MP, a phase with completely filled configuration, CF, and a mixed MP+CF phase. The term 'cluster' is used to denote successively occupied sites. The evolution of clusters in continuous models was described in the review by Mahnke et al \cite{MKL05}.

Various generalizations of the totally asymmetric simple exclusion process (TASEP) have been used to model traffic flow with a special interest in the formation and spreading of traffic jams, see, e.g. the reviews \cite{N96,CSS00,H01}. The jams are known to occur either spontaneously in congested traffic, or due to different inhomogeneities on the road. In this context, the dynamics of the clusters of jammed vehicles is of primary interest. The problem is of interest also for various transport processes taking place in the living cells.
From the viewpoint of physics, molecular motors are proteins or macromolecular complexes that utilize chemical energy to move collectively on a single track in a way resembling vehicular traffic \cite{TC08}. Detailed models of such motor traffic systems describe the mechanochemistry of individual motors with the aim to investigate their collective dynamics \cite{GGNSC07}. Advances of experimental methods have allowed to follow the transport of individual molecules through nanochannels. The effect of crowding of particles in such channels in an experimentally relevant non-equilibrium steady state regime was analyzed in \cite{ZPB09}. A method for analysis of one-dimensional cellular automata models, based on the cluster-size distribution was proposed by Ezaki and Nishinari \cite{EN12}. A comprehensive presentation of traffic flow dynamics and modeling, including microscopic many-particle models, can be found in the book \cite{TK13}.

The model we study here is designed to be simple rather than realistic. In the framework of particle hopping models of traffic, we model jams of vehicles which closely follow the motion of the leading vehicle.

We note that the random-sequential, forward-ordered, sublattice-parallel, and parallel updates of the TASEP configurations, for the definitions see \cite{RSSS98}, violate the integrity of the existing clusters of nearest neighbor particles with nonvanishing probability. The only dynamics that can move forward clusters as a whole entity is the backward-ordered one. However, the probability for translation of a cluster of $k$ particles one site to the right is $p^k$, while such a cluster is broken into two parts with probability $p - p^k$. Thus, in a stationary state of the model on an infinite chain, a specific cluster-size distribution with finite mean cluster size takes place. Remarkably, different types of cluster statistics may appear depending on the dynamic rules and the boundary conditions. For example, exact and asymptotic results on the cluster statistics generated by TASEP on finite rings, with two different dynamics, were obtained in \cite{PM01} by mapping on the zero range process. The results obtained for the standard TASEP show that in the limit of infinite system size, the mean cluster size converges to a finite value, depending on the particle density, while the expectation value of the size of the longest cluster diverges logarithmically with the number of sites on the infinite ring. Since in our model the cluster growth is unlimited on the infinite chain, we consider the stationary cluster size distributions appearing on finite open chains. At that, we focus on the probability $P(1)$ of a completely jammed configuration on segments containing large number of sites $L \gg 1$.

The paper is organized in five sections. In Section 2 we formulate the model, present the phase diagram and those properties which were obtained in \cite{BP16}. In Section 3 we derive some new results about the local density of cars at the ends of the chain. Here the role of the probability P(1) of finding completely jammed configurations on the road segment is elucidated and an exact (in the thermodynamic limit) analytic expression for P(1) is derived. A central result of the paper is the theoretical establishment of three regimes of asymptotic growth of P(1) on approaching the threshold of complete jamming. A discussion of the results and some  perspectives for further applications are given in the concluding Section 4.

\section{The Model}

Actually, our model is a special case of a more general TASEP kinetics. First it was promoted as an exactly solvable generalization of the TASEP on a ring by W\"{o}lki in 2005 \cite{W05}.
We consider it on an open chain of $L$ sites, labeled consecutively by the index $i = 1,2,\dots ,L$.
Each site of the lattice can be empty or occupied by just one particle.

\begin{figure}[t]
\includegraphics[width=100mm]{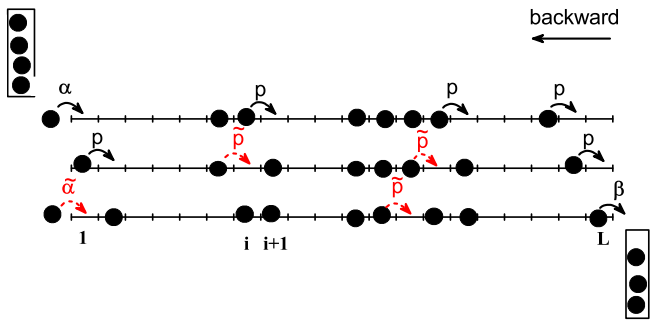}
\vspace{-1cm}
\caption{(Color online) A sketch of the algorithm for update of the gTASEP configurations, illustrated for three consecutive updates. The bulk hopping rules imply that a single particle, as well as a whole cluster of particles, hops one site to the right with probability $p$, provided the target site was vacant. Such a hop is shown by black curved arrow with $p$ above it. Red dashed arrows with $\tilde{\rm p}$ above it illustrate a hop with the modified probability $\tilde{p}$ (see text). Our model corresponds to the limit case of  $\tilde{p} = 1$. The left boundary condition also depends on whether the first site was vacant (black solid arrow with $\alpha$ above it) or it was occupied but became vacant during the backward update (red dashed arrow with $\tilde{\alpha}$ above it).}   \label{Sketch}
\end{figure}

The dynamics of the model corresponds to the discrete-time backward-ordered update with
probabilities $p$ and $\tilde{p}$ defined as follows. A particle
can hop to a vacant nearest-neighbor site in the forward direction, or stay at its place.
During each moment of time $t$, the probability of a
hop along the bond $(i,i+1)$ depends on whether a particle has jumped from site $i+1$ to
site $i+2$ in the previous step, when the bond $(i+1,i+2)$ was updated, or not. The update of the whole
configuration comprises the following steps.

(1) When the last site $L$ is occupied, the particle is removed from it with probability $\beta$, and stays in place with probability $1-\beta$.

(2) Next, the configuration of the system passes through $L-1$ consecutive updates of all the pairs of nearest-neighbor sites in the backward order $(L-1,L), \dots, (i,i+1),\dots, (1,2)$. If site $i+1$ remains empty after the update of $(i+1,i+2)$, then the jump of a particle from site $i$ to site $i+1$ takes place with probability $p$, and the particle stays immobile with probability $1-p$; if site $i+1$ remains occupied, no jump takes place and the configuration of the bond $(i,i+1)$ is conserved. If in the previous step a particle has jumped from site $i+1$ to site $i+2$, thus leaving $i+1$ empty, then the jump of a particle from site $i$ to site $i+1$ in the next step takes place with a different probability $\tilde{p}$, and the particle stays immobile with probability $1-\tilde{p}$.

(3) Finally, the first site is updated by applying the following boundary condition: a particle is injected at the first site of the chain with probability $\alpha >0$,  if the site was vacant at that moment of time, or with probability $\tilde{\alpha} = \min\{\alpha \tilde{p}/p, 1\}$, if the site was initially occupied but became vacant after the update of the bond $(1,2)$ at the same moment of time.

The different possible updates are shown schematically in Fig.~\ref{Sketch} for three consecutive moments of time, along with the corresponding occurrence probability. Note that when $\tilde{p} = p$ one has the standard TASEP with backward-sequential update, and when $\tilde{p} = 0$ one has the TASEP with parallel update. In the general case of $0<\tilde{p}<1$, the model was studied on a ring in \cite{DPPP,PhD,DPP15,AB16}.

We believe that even simple particle hopping models, such as gTASEP, can be useful in describing essential features of real traffic flow \cite{TK13}. For example, the second hopping rate $\tilde{p}$ describes a kind of kinematic attraction ($\tilde{p} > p$), or repulsion ($\tilde{p} < p$), between the vehicles. The case of attraction reflects the natural tendency of the driver to catch up with the car ahead. Thus, clusters of synchronously moving cars appear. Under the gTASEP dynamics, a cluster of $k$ cars is translated  by one site as a whole entity with probability $p\tilde{p}^{k-1}$, and is broken into two parts with probability $p(1-\tilde{p}^{k-1})$. Obviously, the case of stochastic translation of clusters as a whole, combined with their irreversible growth, can be described by the gTASEP in the limit $\tilde{p} =1$. At that, clusters of cars translate by one site forward with the same probability $p$ with which single vehicles move.

A special attention should be paid to the interpretation of the open boundary conditions at the ends of a finite segment of the road. We can suggest one to consider the endpoints of a portion of a single-lane road as toll pay points, operating independently with different efficiency, say, proportional to $\tilde{\alpha}$ at the entrance, and to $\beta$ at the exit of the considered road segment. We remind the reader that the introduction of the $\tilde{p}/p$-dependent injection probability $\tilde{\alpha}$, instead of simply $\alpha$, is necessary for consistency with the special cases of backward-ordered sequential algorithm ($\tilde{p}=p$) and the parallel one
($\tilde{p}=0$). This left boundary condition has been introduced first by Hrab{\'a}k in \cite{PhD}.

\begin{figure}
\vspace{-1cm}
\includegraphics[width=80mm]{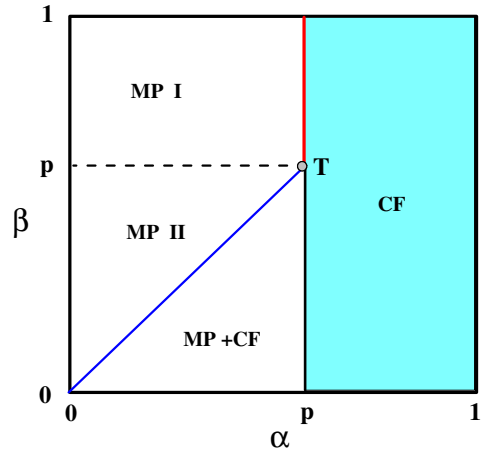} \caption{(Color online) Phase diagram in the plane of injection ($\alpha$) - ejection ($\beta$) probabilities. The many-particle phase MP occupies two regions, MP~I and MP~II; it contains a macroscopic number of single cars or clusters of jammed cars of size $O$(1) as $L\rightarrow \infty$; MP~I and MP~II differ only by the shape of the local density profile. The configurations of the boundary perturbed phase MP+CF contain a macroscopic number of cars and small jams of size $O$(1) close to the left end of the chain, or a single jam  completely filling the whole road. The stationary non-equilibrium phase CF consists of a completely jammed configuration and carries the current $J = \beta$. The small circle with the letter T denotes the triple point. The unusual phase transition, discussed in Ref. \cite{BP16}, takes place across the boundary $\alpha = p$ between the MP~I and CF phases.}   \label{PhaseDiag}
\end{figure}

As shown in \cite{BP16}, the model with irreversible aggregation (jam formation) has three stationary phases: a many-particle one, MP, a phase with completely filled configuration, CF, and a mixed MP+CF phase,
see Fig. \ref{PhaseDiag}.

Note that in the case of $\tilde{p} = 1$, properties (a) and (b) are completely fulfilled. Property (c) is related to the \emph{symmetric} random walk behavior of the gap between two consecutive clusters \emph{in the bulk} of the chain: at each time step it decreases or increases with the same probability $p(1-p)$ and does not change with probability $r = 1-2p(1-p)$. However, when a cluster reaches the last site of the chain, it starts hopping forward with probability $\beta$, instead of $p$. Then the random walk behavior of the gap between this cluster and the one on the left of it changes from symmetric to asymmetric one.

\subsection{Local density}

From the viewpoint of traffic, it is important to know the magnitude of the flow, and the car density both at the entrance and exit of the road. In the completely jammed phase CF, the car density profile is completely flat, $\rho_i \equiv 1$, $i = 1,2,\dots,L$, and the current depends trivially on the ejection probability, $J = \beta$. The local densities at the ends of the road were derived in \cite{BP16} only for the MP phase ($\alpha <p$ and $\beta >\alpha$), where the profile is flat from the first site up to the bulk, $\rho_1^{\rm MP} = \rho_{\rm b}^{\rm MP}$. From the local balance of the average inflow and outflow of particles at the first site of the chain, the result $\rho_1^{\rm MP} =  \alpha/p$ was obtained; by using the global stationarity condition it was found in addition that $\rho_L^{\rm MP} = \alpha/\beta$. Thus, on the line $\beta = p$ between the subregions MPI and MPII the density profile is completely flat; it bends downwards in MPI, where $\beta >p$, and upwards in MPII, where the opposite inequality $\beta < p$ holds.

In the phase MP+CF ($\beta < \alpha < p$) the road is completely jammed from the bulk up to the last site,  $\rho_{\rm b}^{\rm MP+CF} =\rho_L^{\rm MP+CF} =1$. The particle current is given by $J = \beta$, as in the pure phase CF. However, the road near the entrance is not completely jammed and the local density $\rho_1^{\rm MP+CF}$ decreases with increasing $\beta$ down to $\rho_1^{\rm MP+CF}=\alpha/p$ at $\beta =\alpha$. Here we prove the result
\begin{equation}
\rho_1^{\rm MP+CF}=1 - (1/\alpha -1/p)\beta,
\label{ro1}
\end{equation}
which completes the description of the local density profiles.

The derivation of Eq. (\ref{ro1}), despite its simplicity, needs some reflection. Consider the input-output balance of particles at the first site. Clearly, the probability that the empty site $i=1$ will become occupied by a particle is $\alpha (1- \rho_1)$. The problem is that the probability with which the occupied first site will become empty at the end of an update depends on whether the configuration is completely jammed, or not. In the former case, which takes place with probability P(1), the particle at the first site belongs to the cluster occupying all the sites in the chain, including the exit one $i=L$. Hence, the site $i=1$ will become vacant with probability $\beta$, when the whole cluster moves one site to the right, and will remain vacant with probability $(1-\tilde{\alpha})$ at the end of that update. On the other hand, with probability 1 - P(1) the particle at the first site is isolated, or belongs to a cluster which is separated by vacant site(s) from the end of the chain. In this case, the first site becomes vacant with probability $p$, with which the cluster it belongs to moves one site to the right, and remains vacant again with probability $(1-\tilde{\alpha})$. Thus, the particle balance at the first site cannot be satisfied during each single configuration update.

To solve the problem, we take into account that the stationary current $J_{i,i+1}$ through a bond $(i,i+1)$ is constant along the chain, $J_{1,2}= J_{2,3}=\dots = J_{L-1,L} = J_{\rm out}$, where $J_{\rm out} =\beta \rho_L$
is the current of particles leaving the chain. Since in our case $\rho_L =1$, the global stationarity of the occupation number at the first site requires the equality  $\alpha (1- \rho_1)= (1-\tilde{\alpha})\beta$, which implies the result (\ref{ro1}).

\subsection{Probability of complete jamming}

An expression for the probability P(1) of completely jammed configurations of the system in the mixed MP+CF phase can be derived from the global stationarity condition $J_{\rm in} = J_{\rm out}$, where
$J_{\rm in}$ ($J_{\rm out}$) is the input (output) current. A particle may enter the system in two cases: (i) If the system is in a completely jammed configuration, which happens with probability P(1), the first site becomes empty with probability $\beta$, as a result a particle ejection from the last site of the chain and a deterministic shift of the whole cluster one site to the right; then a particle may enter the system with probability $\tilde{\alpha} = \alpha/p$. The total probability of this event is P(1)$(\alpha/p)\beta$. (ii) When the system is not completely jammed, which takes place with probability 1-P(1), a particle enters at the first site with probability $\alpha$, as it was proven in \cite{BP16}.
By equating the input current
\begin{equation}
J_{\rm in}= P(1)\frac{\alpha \beta}{p} + [1-P(1)]\alpha
\label{inputprob}
\end{equation}
to the output current $J_{\rm out}= \beta$ (we recall that $\rho_L^{\rm MP+CF} =1$), one solves for P(1) and obtains:
\begin{equation}
P(1) = \frac{p(\alpha - \beta)}{\alpha (p-\beta)}, \qquad \beta \le \alpha <p.
\label{P1Rational}
\end{equation}

\begin{figure}
\includegraphics[width=100mm]{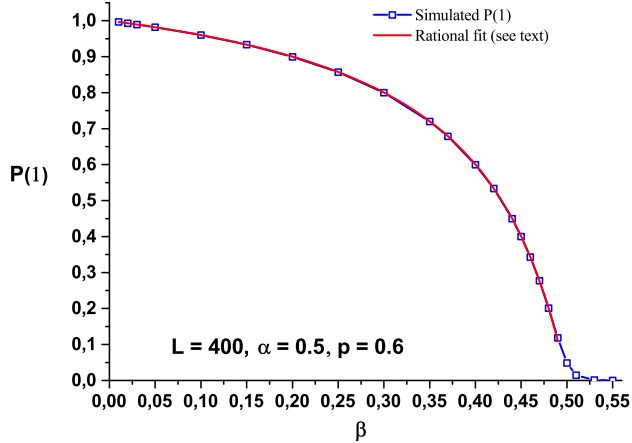} \caption{(Color online) Comparison of the analytic result (\ref{P1Rational}), solid red line, as a function of the ejection probability $\beta$, to the computer simulation, shown by empty blue squares, for a lattice of size $L=400$, injection probability $\alpha = 0.5$ and hopping probability $p = 0.6$.
\label{P1betaFig}}
\end{figure}

This result is in excellent agreement with the Monte Carlo simulation data, see Fig. \ref{P1betaFig}.

Some details of our Monte Carlo simulation techniques are given in the next subsection.

\subsection{Details of the Monte Carlo simulations}

The configuration update rules, described at the beginning of the section, were implemented in a code for Monte Carlo simulations, written in Fortran 90. The code was ran on a personal computer with an Intel Core i5-3350P 3.1 GHz processor and 8 GB RAM. The standard generator of pseudo-random numbers ran2 was used \cite{NRec96}.

Typically, each data point was obtained by averaging over 100 independent temporal series (runs) of length $4\times 10^6$ configuration updates for chains of length 400 sites. In every time run, $2\times 10^6$ lattice updates were omitted before the data collection starts. Beforehand, tests were made, starting from different initial configurations (e.g., a completely empty lattice or with a given initial density), and averaging over time series of different length. The time behavior of the bulk density was recorded, and the number of runs, their length and the number of omitted updates between them, were chosen as to ensure that the system reaches a stationary state with acceptable accuracy. Stationarity was controlled also by monitoring the input and output currents: they were allowed to differ by less than $10^{-5}$ in absolute units.

To assess the finite-size effects we have studied lattice sizes from $L= 200$ to $L= 1200$. Optimal balance between CPU time and relative accuracy within $10^{-3}$ for local particle densities, and $10^{-4}$ for global ones (bulk density and current) was achieved for chains with $L=400$ sites.

\section{Random walk theory}

\subsection{Evolution of the system configurations}

\begin{figure}[t]
\begin{minipage}{55mm}
\center
\includegraphics[width=40mm]{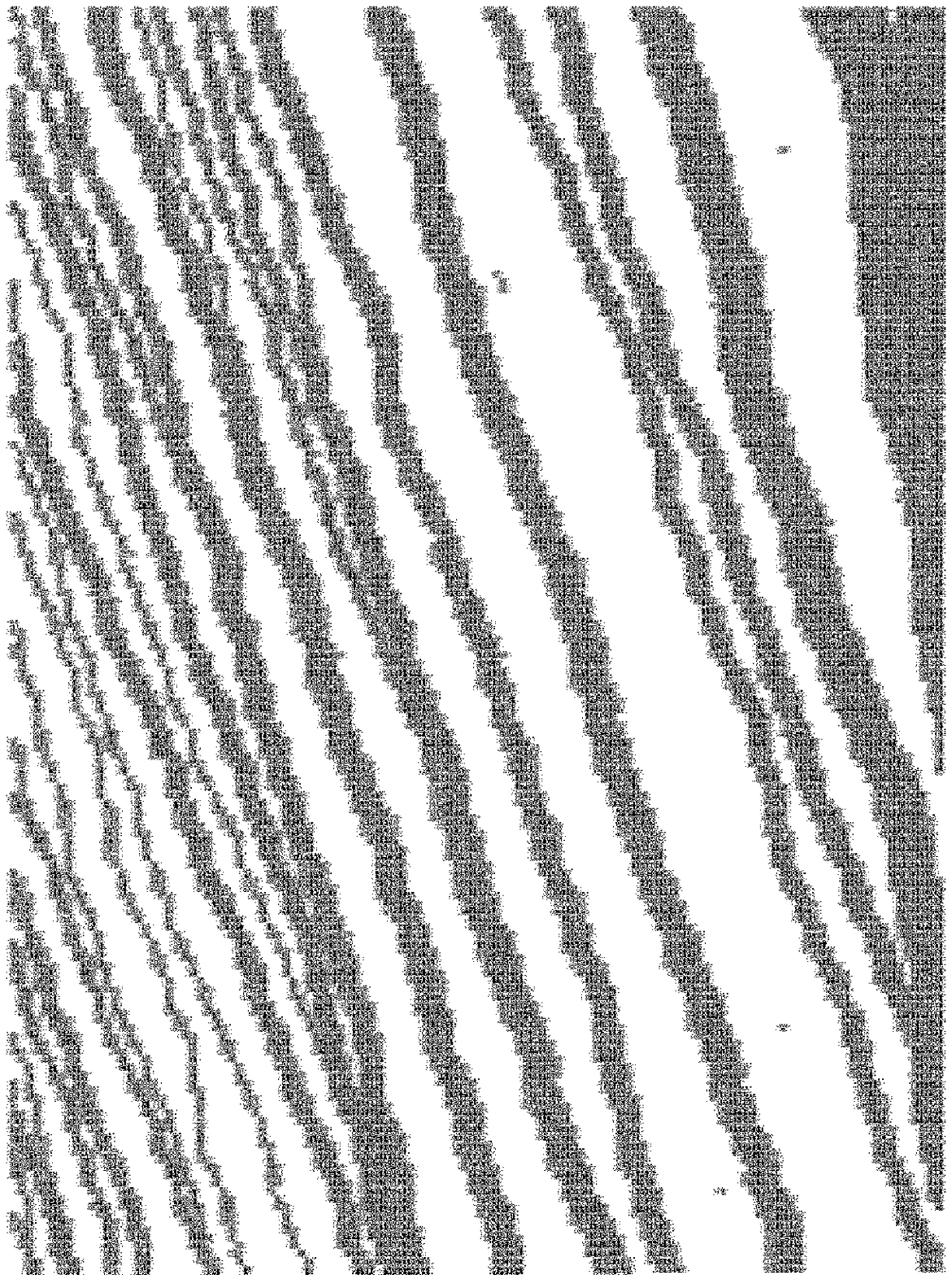}
\endcenter
\caption{\label{Fig4} Simulated time evolution of clusters and decreasing gaps (white strips) in the MPII phase, at $L=200$, $\alpha =0.2$, $\beta =0.3$, and $p=0.6$.}
\end{minipage}\hspace{2pc}
\begin{minipage}{55mm}
\center
\includegraphics[width=40mm]{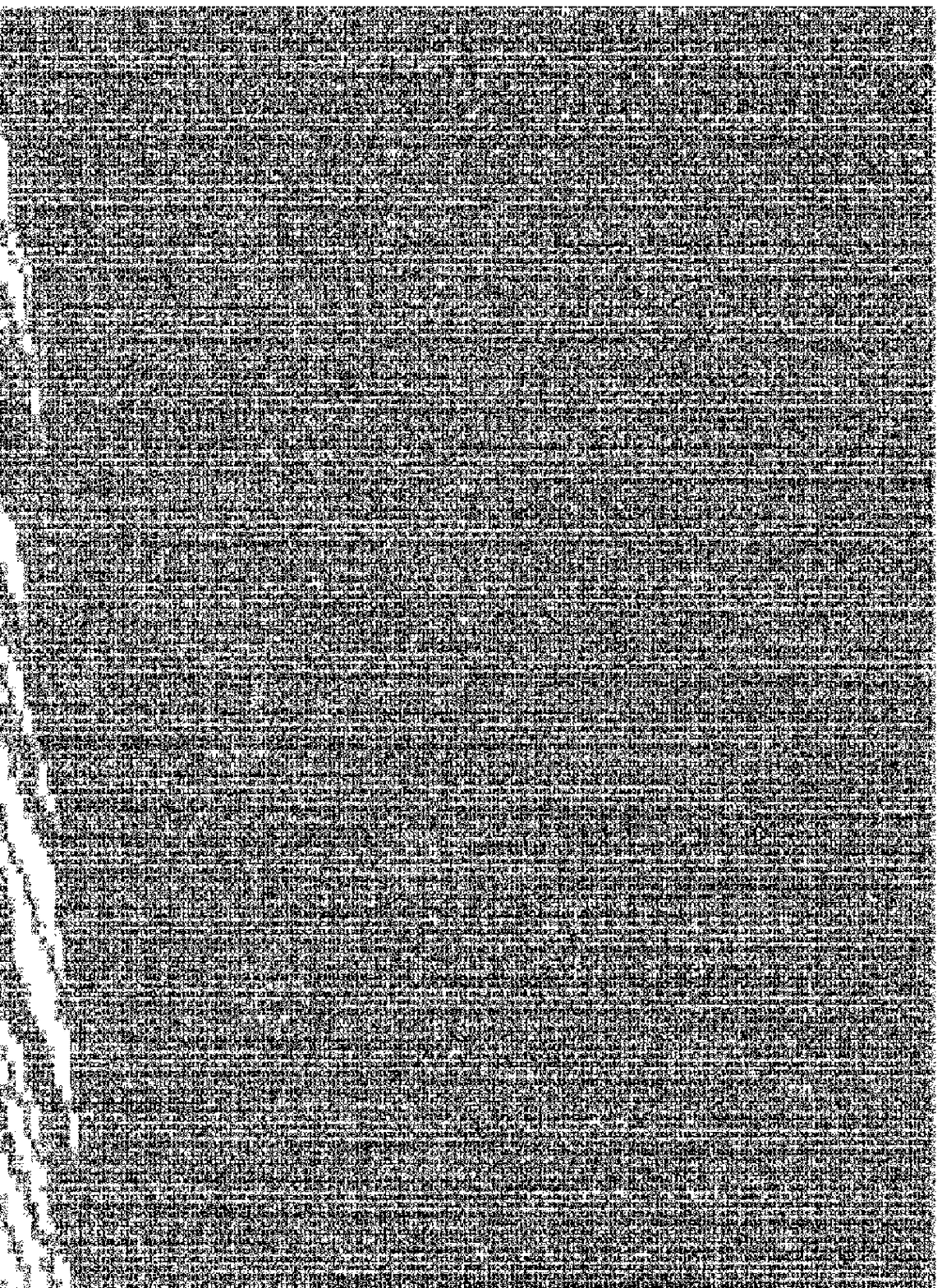}
\endcenter
\caption{\label{Fig5} Simulated time evolution of clusters and gaps (white strips), in the boundary perturbed MP+CF phase, at $L=200$, $\alpha =0.31$, $\beta =0.3$, and $p=0.6$.}
\end{minipage}
\end{figure}

In the particle hopping discrete-space traffic models, one often uses the dual representation of configurations in terms of empty sites positions, instead of particle coordinates. In the case of our model, such a representation leads to a very peculiar dynamics of the inter-cluster gaps. First, because existing clusters cannot break up, gaps may appear only at the first site of the chain. Second, gaps disappear when two consecutive clusters merge or when the rightmost cluster leaves the system and the one following it reaches the last site of the chain. Third, and most interesting feature of the gap dynamics is that, as long as two consecutive gaps exist, the distance between them remains constant. Indeed, because of particle conservation in the bulk, the number of particles in a cluster between two gaps remains fixed. Thus, in the space-time picture, the neighboring edges of two coexisting gaps are forced to move in parallel, see a computer illustration in Fig.~\ref{Fig4}, and gaps may not cross or merge. Finally, the width of each gap performs a random walk, which begins with an initial state containing one or several neighboring empty sites, and ends up when the random walk reaches to the origin. If a configuration has two or more gaps, the random walks performed by their edges are not independent.


\medskip

In general, when $\beta \not= p$, the size of the rightmost gap performs an \emph{asymmetric} random walk: its right edge belongs to the cluster which extends to the exit site $i=L$, hence it moves one site to the right with the ejection probability $\beta$ and stays in place with probability $1-\beta$; on the other hand, its left edge moves one site to the right with probability $p$, or stays in place with probability $1-p$. Therefore, after each update, the gap width increases by one site with probability $p_g$, decreases by one site with probability $q_g$, and remains the same with probability $r$, where
\begin{equation}
p_g = \beta (1-p),\quad q_g = p(1-\beta), \quad r = 1 - q_g - p_g = 1-\beta -p +2\beta p.
\label{assRW}
\end{equation}

In contrast, the size of all gaps on the left of the rightmost one performs a \emph{symmetric} random walk: both edges of each of these gaps execute a directed random walk with hopping probability $p$. Hence, for the corresponding gap width one has
\begin{equation}
p_g = q_g =p(1-p),\quad r = 1- 2p(1-p).
\label{symRW}
\end{equation}
However, since the inner edges of all pairs of coexisting consecutive gaps are forced to move synchronously, we have a set of \emph{interacting random walks}: independent is only the random walk of the external edges: the right edge of the rightmost gap and the left edge of the leftmost gap.

We note that there is a close relationship between the asymmetry of the random walk, performed by the width of the rightmost gap, and the shape of the stationary local density profile in the MP phase. When $\beta >p$
($\beta < p$) and the width of that gap increases (decreases), the stationary local density profile bends down (up) close to the chain end, see Fig. 3 in Ref. \cite{BP16}.

A considerable simplification of this picture takes place when $\beta =p$: the widths of all the existing gaps perform symmetric random walks, starting from the corresponding initial conditions. In this case, the local density profile is completely flat.

In principle, the lifetime probability distribution for all the gaps can be calculated, provided the initial conditions are given. However, the problem of calculating the stationary probability P(1) of a completely filled configuration by averaging over the time evolution of the system remains hardly solvable.

\subsection{Evolution of single-gapped configurations}

We can estimate analytically the probability P(1) of a completely filled configuration in the case when the appearance of a vacancy at the first site of a completely filled configuration is a rare event. That event happens when: (i) a particle of the completely filling cluster leaves the system from its right end, which takes place with probability $\beta$, and leads to the deterministic translation of all the remaining $L-1$ particles by one site to the right; (ii) the resulting vacancy of the first site is not immediately filled by a particle from the left reservoir, which takes place with probability $(1-\tilde{\alpha})\ll 1$. Thus, the appearance of a vacancy at the first site of a completely filled configuration will take, on the average,
\begin{equation}
\bar{N} = [(1-\tilde{\alpha})\beta]^{-1}\gg 1
\label{Nc}
\end{equation}
updates (discrete time steps).

In the course of time, the single vacancy may give rise to gaps of size one or more empty sites between the clusters of filled sites.
Thus, we have to evaluate the average lifetime (in number of updates) $\bar{n}$ of the gap in the different asymptotic regimes. If $\bar{n}< \bar{N}$, an estimate of the probability $P(1)$ is given by the ratio
\begin{equation}
P(1) \simeq (\bar{N} - \bar{n})/\bar{N}.
\label{P1est}
\end{equation}

\subsection{Growing gap}

First, consider the case when the boundary $\alpha = p$ of the CF phase is approached from region MPI, i.e., at $\beta >p$. In this case $p_g > q_g$ and the gap asymptotically grows with the number $n\gg 1$ of time steps as $W(n) \propto (p_g - q_g)n$. The gap will exist until its lower edge reaches the end of the chain, that is for $L/p$ time steps on the average. The two edges of the gap coexist until the time moment $L/\beta$ when the upper edge reaches the end of the chain. According to the asymmetric random walk
theory, the probability that a gap with unit initial width collapses, before the time moment $L/\beta$, tends to $q_g/p_g < 1$, as $L\rightarrow \infty$. Therefore, the probability that the gap, which has opened at site $i=1$, survives until the end of the chain is
\begin{equation}
P_{surv}= 1-q_g/p_g =\frac{\beta-p}{\beta(1-p)}
\label{surv}
\end{equation}
Then, a long gap, propagating through the whole chain, appears on the average after
\begin{equation}
\bar{N_c}=\frac{\bar{N}}{P_{surv}}=\frac{p(1-p)}{(p-\alpha)(\beta-p)}
\label{long}
\end{equation}
updates. Hence, expression (\ref{P1est}) becomes
\begin{equation}
P(1) \simeq (\bar{N_c} - \bar{n})/\bar{N_c}.
\label{P1ex}
\end{equation}.

\begin{figure}[t]
\includegraphics[width=110mm]{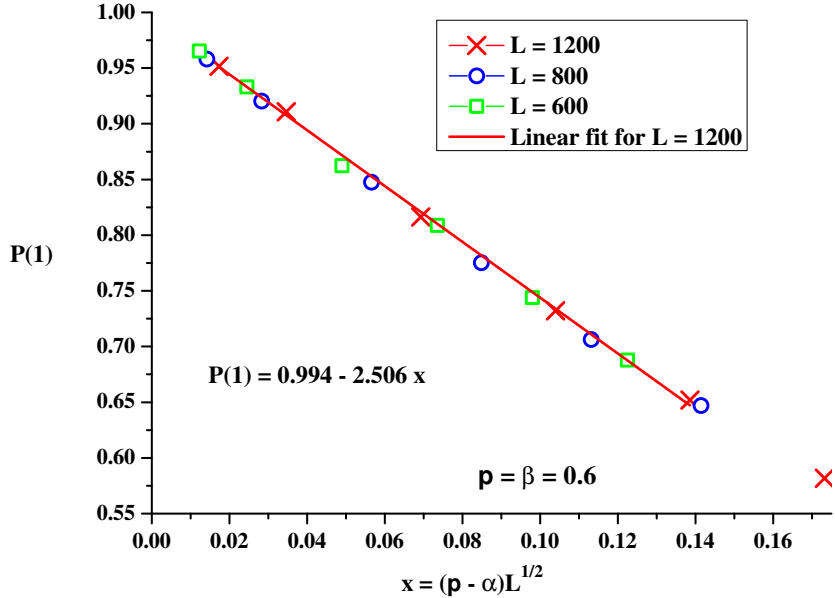} \caption{(Color online) Data collapse of the computer simulation data for the probability P(1) of completely jammed configuration at $\beta = 0.9$, as a function of the finite-size scaling variable $x=L(p-\alpha)$ for different chain lengths $L$. The linear fit to the small $x$ asymptotic behavior for $L=800$ is shown by solid red line.}\label{CollP109}
\end{figure}

The crucial assumption is that $\bar{n} = L/p \ll \bar{N_c}$, and no other long-living gap appears during the considered time interval $\bar{N_c}$. In such a case, from Eqs. (\ref{long}) and (\ref{P1ex}) we obtain the result,
\begin{equation}
P(1) \simeq 1 - a_1 L(p-\alpha),
\label{P1upper}
\end{equation}
where
\begin{equation}
a_1 = \frac{\beta-p}{p^2(1-p)}
\label{a1}
\end{equation}
The important result here is the appearance of the finite-size scaling variable $L(p-\alpha)$, which is confirmed by applying the data collapse method to the results of our computer simulations for different chain lengths $L$, see Fig. \ref{CollP109}. Inserting $\beta = 0.9$ and $p = 0.6$ into (\ref{a1}) we obtain $a_1\simeq 2.08$, which agrees very well with $a_1^{\rm sim} \simeq 2.063$ evaluated from Fig. \ref{CollP109} for $L=400$.

Note that the neglect of short-living gaps implies that Eqs. (\ref{P1upper}) and (\ref{a1}) actually give an upper bound for P(1). The lower bound corresponds to $P_{surv}= 1$, which leads to $a_1 = \beta/p^2$.

Next, using the stationary condition $J_{\rm in}=J_{\rm out}$ and the value of $J_{\rm out}=\beta \rho_L$ at the end of the chain, we obtain from Eq. (\ref{inputprob}), in the vicinity of $\alpha = p$,
\begin{equation}
\rho_L = P(1)+ [1-P(1)]\frac{\alpha}{\beta}
\label{L}
\end{equation}
and
\begin{equation}
J=P(1)\beta + [1-P(1)]\alpha
\label{J}
\end{equation}

\begin{figure}
\includegraphics[width=110mm]{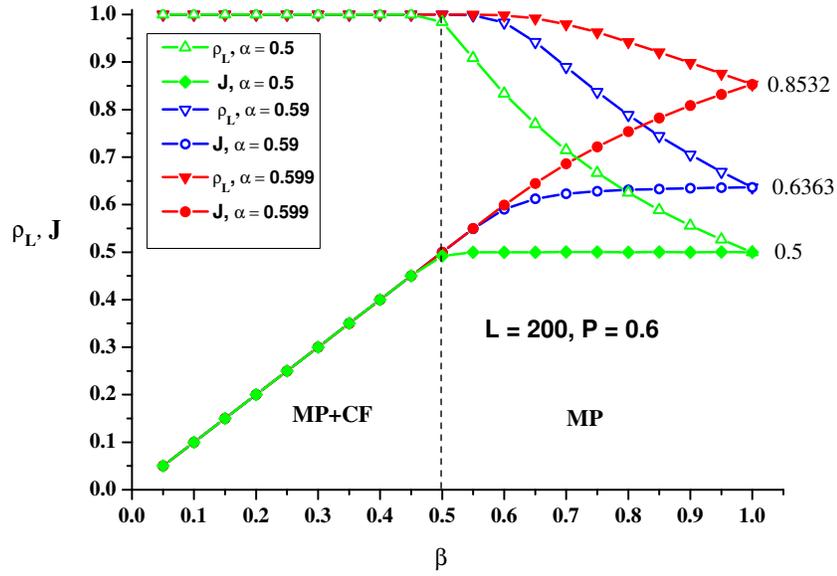} \caption{(Color online) Illustration of the change in the behavior of the current and the local particle density $\rho_L$ in the MP phase, when $\alpha$ increases, approaching the boundary $\alpha = p$ with the CF phase. Computer simulation data for $L=200$, $p=0.6$ and three values of $\alpha$ are shown. At $\alpha = 0.5$ one observes the usual first-order phase transition that takes place at  $\beta = \alpha <p$.  The unusual phase transition, discussed in Ref. \cite{BP16}, takes place across the boundary $\alpha = p$ between the MP and CF phases.}   \label{RoJPout}
\end{figure}

Expressions (\ref{L}) and (\ref{J}) allow us to explain the so called 'unusual' phase transition between the MP and CF phases, found in Ref. \cite{BP16}. We recall that this non-equilibrium 'zeroth-order' phase transition at $\beta > p$ manifests itself by jumps both in the density $\rho_L(\alpha)$ and the current $J(\alpha)$ at the boundary $\alpha=p$, $\beta>p$, between the MPI and CF phases, see Fig.~7 in \cite{BP16}. This transition appears instead of the usual non-equilibrium second-order transition between the low-density (high-density) and the maximal current phases in the standard TASEP.
The MPI phase is characterized, in general, by vanishing probability P(1) of completely jammed configurations. However, close to the boundary $\alpha = p$, $\beta>p$, the value of P(1) increases rapidly as $(p-\alpha)\rightarrow 0$ and the jumps (in the limit $L\rightarrow \infty$) in the $\alpha$-dependence of $\rho_L(\alpha)$ and $J(\alpha)$ are a direct consequence of Eqs. (\ref{L}) and (\ref{J}). The effect of the growing P(1) on the current and local particle density $\rho_L$, when $\alpha$ approaches the boundary between the MP and CF phases, is illustrated in Fig. \ref{RoJPout}.

\subsection{Critical gap}

When the boundary $\alpha = p$ of the CF phase is approached along the line $\beta = p$, from Eq. (\ref{assRW}) we obtain that the gap width performs a symmetric random walk with $p_g = q_g = p(1-p)$, $r = 1 -2p(1-p)$.
The calculation of $\bar{n}$ in this case becomes involved, since the average length of a symmetric random walk on the infinite chain, with initial state at $i=1$ and one absorbing state at $i=0$, diverges. That is why, we have to take into account that the lifetime of the symmetric random walk on the finite chain is limited, on the average, by $L/p$. Let us introduce the average lifetime $\bar{n}_M$ of the gap during $M$ updates,
\begin{equation}
\bar{n}_M = \sum_{m=1}^M m f_{1,0}^{(m)},
\end{equation}
where $f_{1,0}^{(m)}$ is the probability that the initial gap of unit size will vanish at update $m$. By
using the generating function of $f_{1,0}^{(m)}$, see Eq. (53) in Ref. \cite{CM01},

\begin{figure}
\includegraphics[width=110mm]{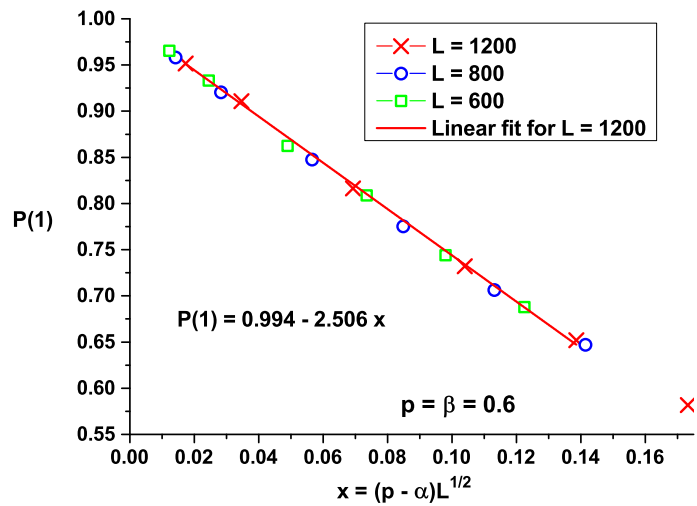} \caption{(Color online) Data collapse of the computer simulation data for the probability P(1) of completely jammed configuration at $\beta = p = 0.6$, as a function of the finite-size scaling variable $x=L^{1/2}(p-\alpha)$ for different chain lengths $L$. The linear fit to the small $x$ asymptotic behavior for $L=1200$ is shown by solid red line.}
\label{CritGapColl}
\end{figure}

\begin{equation}
F_{1,0}(s) = \sum_{m}f_{1,0}^{(m)}s^m = \left\{1-r s - \left[(1-r s)^2 -4 p^2(1-p)^2 s^2\right]^{1/2}\right\}/[2p(1-p)s],
\end{equation}
we obtain the exact result
\begin{equation}
\bar{n}_M = \frac{p(1-p)}{r}\sum_{n=0}^{M/2-1}\left(\frac{p(1-p)}{r}\right)^{2n} \left(\begin{array}{c}  2n+1 \\ n \end{array}\right)
\sum_{k=2n+1}^M r^k \left(\begin{array}{c}  k \\ 2n+1 \end{array}\right).
\end{equation}
In the case of $M\gg 1$, we use the approximation
\begin{equation}
\sum_{k=2n+1}^M r^k \left(\begin{array}{c}  k \\ 2n+1 \end{array}\right)\simeq \sum_{k=2n+1}^{\infty} r^k \left(\begin{array}{c}  k \\ 2n+1 \end{array}\right) = \frac{r^{2n+1}}{(1-r)^{2n+2}},
\end{equation}
which leads to
\begin{equation}
\bar{n}_M \simeq \frac{1}{4p(1-p)}\sum_{n=0}^{M/2}\frac{1}{4^{n}} \left(\begin{array}{c}  2n+1 \\ n \end{array}\right)\simeq \frac{1}{2\sqrt{\pi}p(1-p)}\int_0^{M/2} \frac{{\rm d} x}{\sqrt{x}} = \frac{\sqrt{M}}{\sqrt{2\pi}p(1-p)}.
\end{equation}
Finally, by setting $M = L/p$, from Eqs. (\ref{Nc})  and (\ref{P1est}) at $\beta =p$, we obtain
\begin{equation}
P(1) \simeq 1 - b_1 L^{1/2}(p-\alpha), \qquad b_1 = \frac{1}{\sqrt{2\pi p}p(1-p)}.
\label{P1Crit}
\end{equation}
The important feature here is the appearance of the critical finite-size scaling variable $x=L^{1/2}(p-\alpha)$, which is confirmed by the results of our computer simulations, see Fig. \ref{CritGapColl}. In spite of the crude approximations, the estimate of the constant $b_1$ is rather good: at $p = 0.6$ Eq. (\ref{P1Crit}) yields $b_1 \simeq 2.15$, while the corresponding values obtained from the small $x$ asymptotic behavior shown in Fig. \ref{CritGapColl} is $b_1^{\rm sim} \simeq  2.51$.

\subsection{Short-living gap}

When the boundary of the CF phase is approached from the mixed MP+CF phase, i.e., at $\beta < p$, one has $p_g < q_g$ and the gap between the newly growing cluster and the cluster which is leaving the system from its right boundary closes in a finite number of time steps. An upper estimate can be given by the result for a random walk on an infinite chain with initial state at site $i=1$ and one absorbing state at the origin $i=0$:
\begin{equation}
\bar{n} <  (q_g - p_g)^{-1} = (p-\beta)^{-1}.
\label{P1finite}
\end{equation}
Obviously, in this case $P(1) \rightarrow 1$ as $\alpha \rightarrow p$ at $\beta < p$.

\section{Discussion}

We have reformulated the one-dimensional model of irreversible aggregation of particles, proposed in Ref. \cite{BP16}, as a model of irreversible jam formation in dense traffic on finite segments of a single-lane road. In the above reference, the stationary state values of the local density at the endpoints of the road segment were derived only in the MP phase. Here we have completed this study by deriving an exact expression for the stationary density at the first site of the chain in the mixed MP+CF phase, see Eq. (\ref{ro1}). From the bulk up to the last site, the road segment is completely jammed, $\rho_{\rm b}^{\rm MP+CF} =\rho_L^{\rm MP+CF} =1$. We have also derived an exact expression for the probability P(1) of complete jamming in the mixed MP+CF phase, see Eq. (\ref{P1Rational}) and Fig. \ref{P1betaFig}. The appearance of the unusual 'zeroth order' phase transition, found in Ref. \cite{BP16}, has been explained by the rapid increase of the P(1) contribution to both the local density $\rho_L$ and the current $J$, see Eqs. (\ref{L}) and (\ref{J}), as the particle injection probability $\alpha$ approaches from the left the boundary $\alpha =p$, $\beta >p$, between the MP and CF phases.

In addition, we have analyzed, within the random walk theory, the evolution of a single gap between two large clusters of vehicles. Three qualitatively different regimes were established when the injection rate approaches from the left the boundary $\alpha =p$ with the CF phase: (i) the lifetime of the rightmost gap in the jammed configuration is of size $O(L)$ in the MP phase; (ii) macroscopic jams, separated by short-living gaps of length $O(1)$, exist in the MP+CF phase; and (iii) there is a critical regime when the macroscopic jams are divided by gaps of intermediate lifetime of the order $O(L^{1/2})$ when the triple point $\alpha =\beta =p$ is approached. These results are supported by extensive Monte Carlo calculations.

The microstructure of traffic flow is usually described in terms of the probability distribution of the distance gap and time-headway, see a survey in \cite{LC11}. For example, the exact analytical expression for the probability distribution of the distance headways in the steady state of Nagel-Schreckenberg model of vehicular traffic at $V_{\rm max} = 1$ was obtained by Chowdhury et al \cite{CMGSS97}. The time- and distance-headways in the general case of that model were studied in Ref. \cite{CPS98}. A microscopic theory of spatial-temporal congested traffic patterns in heterogeneous traffic flow with a variety of driver behavioral characteristics and vehicle parameters was presented by Kerner and Klenov in \cite{KK04}. The time-headway distribution for the TASEP under various updates, including the generalized one, but under periodic boundary conditions only, was reported by Hrab\'{a}k and Krb\'{a}lek in Ref. \cite{HK16}. However, their results for the gTASEP break down in our limiting case of $\tilde{p} =1$ ($p\gamma =1$ in their notation). We are not aware of analogous results for open chains. Surprisingly, the authors of Ref. \cite{KS15} demonstrated that the vehicular gap distributions in the vicinity of a signalized intersection are a consequence of the general stochastic nature of queueing systems, rather than a consequence of traffic rules and drivers behavior.

One may speculate about the usefulness of the gap representation of the system configurations, as compared to the cluster one. In a generic point in the MP phase, the two representations seem equivalently complicated, see Fig. \ref{Fig4}.  Nevertheless, the gap representation seems promising, because one has an ensemble of equivalent critical gaps, besides the growing rightmost one (when $\beta >p$), which start at the left boundary, do not merge or cross and their probability distribution is known. In a forthcoming paper we are going to use the gap presentation for calculation of the density profile over the whole chain interval $[0,L]$.

\section*{Acknowledgement} The partial support by a grant of the Plenipotentiary Representative of the Bulgarian Government at the Joint Institute for Nuclear Research, Dubna, is gratefully acknowledged. VBP acknowledges the support of the RFBR, grant 16-02-00252.

\end{document}